\documentclass[11pt,a4]{llncs}

\usepackage[T1]{fontenc}
\usepackage[utf8]{inputenc}
\usepackage[british]{babel}
\usepackage{amssymb,amsmath}
\usepackage{times}
\usepackage{xspace}

\usepackage{url}
\usepackage{colonequals}
\usepackage{listings}

\newcommand{\x}{\xspace}
\newcommand{\coq}{\textsc{Coq}\x}
\newcommand{\setQ}{\ensuremath{\mathbb{Q}}\x}
\newcommand{\setN}{\ensuremath{\mathbb{N}}\x}
\newcommand{\setR}{\ensuremath{\mathbb{R}}\x}
\newcommand{\gp}{\ensuremath{\texttt{p}\x}}

\newcommand{\robid}{\ensuremath{\textsl{r-id}\x}}
\newcommand{\sem}[1]{\ensuremath{[\![#1]\!]\x}}
\renewcommand{\epsilon}{\varepsilon}
\newtheorem{thm}{Theorem}

\usepackage{lstcoq}
\begin{document}

\title{Impossibility of Gathering, a Certification}
 \institute{ 
   {\'Ecole Nat. Sup. d'Informatique pour l'Industrie
     et l'Entreprise (ENSIIE), \'Evry, F-91025}
   \and {\textsc{C\'edric} -- Conservatoire national des arts et
     m\'etiers, Paris, F-75141}
   \and {LRI, CNRS UMR 8623, Universit\'e Paris-Sud, Orsay, F-91405}
   \and {UPMC Sorbonne Universit\'{e}s}
 \and Institut Universitaire de France
   }
\author{Pierre Courtieu\inst{2}
 \and Lionel Rieg\inst{1,2}
 \and Sébastien Tixeuil\inst{4,5} 
 \and Xavier Urbain\inst{1,2,3} 
}

\maketitle

\begin{abstract}
Recent advances in Distributed Computing highlight models and
algorithms for autonomous swarms of mobile robots that self-organise
and cooperate to solve global objectives. The overwhelming majority of
works so far considers handmade algorithms and proofs of correctness. 

This paper builds upon a previously proposed formal framework to
certify the correctness of impossibility results regarding distributed
algorithms that are dedicated to autonomous mobile robots evolving in
a continuous space. As a case study, we consider the problem of
gathering all robots at a particular location, not known beforehand. A
fundamental (but not yet formally certified) result, due to Suzuki and
Yamashita, states that this simple task is impossible for two robots
executing deterministic code and initially located at distinct
positions. Not only do we obtain a certified proof of the original
impossibility result, we also get the more general impossibility of
gathering with an even number of robots, when any two robots are
possibly initially at the same exact location.
\end{abstract}

\section{Introduction}

The Distributed Computing community, motivated by the variety of tasks
that can be performed by autonomous robots and their complexity,
started recently to propose formal models for these systems, and to
design and prove protocols in these models. The seminal paper by
Suzuki \& Yamashita~\cite{suzuki99siam} proposes a robot model, two
execution models, and several algorithms (with associated correctness
proofs) for gathering and scattering a set of robots. In their model,
robots are identical and anonymous (they execute the same algorithm
and they cannot be distinguished using their appearance), robots are
oblivious (they have no memory of their past actions) and they have
neither a common sense of direction, nor a common handedness
(chirality). Furthermore, robots do not communicate in an explicit
way. They have however the ability to sense the environment and see
the position of the other robots. Also, robots execute three-phase
cycles: \textit{Look}, \textit{Compute} and \textit{Move}. During the
\textit{Look} phase, robots take a snapshot of the other robots'
positions. The collected information is used in the \textit{Compute}
phase in which robots decide to move or to stay idle. In the
\textit{Move} phase, robots may move to a new location computed in the
previous phase. The two execution models are denoted (using recent
taxonomy~\cite{flocchini12book}) FSYNC, for fully synchronous, and
SSYNC, for semi-synchronous. In the SSYNC model, an arbitrary
non-empty subset of robots execute the three phases synchronously and
atomically. In the FSYNC model, all robots execute the three phases
synchronously.

One of the benchmarking~\cite{flocchini12book} problems for mobile
robots is that of \emph{Gathering}. Regardless of their initial
positions, robots have to move in such a way that they 
eventually stand 
 on the same location, not known beforehand, and remain there thereafter. A key impossibility result for gathering is due to Suzuki \& Yamashita~\cite{suzuki99siam}: two robots initially located at distinct positions may never gather if they execute a deterministic algorithm. This result is fundamental because any weakening of the initial system hypotheses (\emph{e.g.} anonymity, obliviousness, common sense of direction) makes the problem solvable~\cite{CFPS12j}.

\paragraph{Related Works}

Most related to our concern are recent approaches to mechanising the
algorithm design or the proof of correctness in the context of
autonomous mobile
robots~\cite{bonnet12sss,devismes12sss,BMPTT13r,auger13sss}. Model-checking
proved useful to find bugs in existing literature~\cite{BMPTT13r} and
assess formally published algorithms~\cite{devismes12sss,BMPTT13r}, in
a simpler setting where robots evolve in a \emph{discrete space} where
the number of possible positions is finite. 
However, no method exists to derive impossibility results using model
checking. Automatic program synthesis (for the problem of perpetual
exclusive exploration in a ring-shaped discrete space) is due to
Bonnet \emph{et al.}~\cite{bonnet12sss}, and could be used to prove
impossibility in a particular setting (by a side effect, if no
algorithm can be generated), yet it exhibits important limitations for
studying the gathering problem we focus on here. First, the authors consider
only the discrete space setting (with a ring shape). Second, their
approach is brute force (it generates every possible algorithm in a
particular setting, regardless of the problem to solve). Third, the
generator is limited to configurations where \emph{(i)} a location can
only host one robot (so, gathering cannot be expressed), and
\emph{(ii)} no symmetry appears (which eludes all interesting cases
for studying gathering).

Developed for the \coq proof
assistant\footnote{\url{http://coq.inria.fr}}, the Pactole framework enabled the use of high-order logic to certify
impossibility results~\cite{auger13sss} for the problem of
convergence: for any positive $\epsilon$, robots are required to reach
locations that are at most $\epsilon$ apart. Of course, an algorithm
that solves gathering also solves convergence, but the converse is not
true. As convergence is solvable in the usual setting, the
impossibility results that can be obtained involve Byzantine robots
(that is, robots that may exhibit arbitrary, and possibly malicious,
behaviours). The impossibility results obtained in previous work using
Coq~\cite{auger13sss} show that convergence is impossible if more than
half of the robots are Byzantine in the FSYNC model (or more that one
third of the robots are Byzantine in the SSYNC model). These results
cannot be directly extended to that of Gathering Impossibility for
several reasons. First, they involve the active participation of
Byzantine robots to destabilise the correct ones, while the gathering
problem involves only correct robots. Second, the possible positions
robots may occupy are encoded using rational numbers, while positions
in the original model actually use real numbers.

\paragraph{Our Contribution}

In this paper, we consider the construction of a formal proof for the
fundamental impossibility result of Suzuki and
Yamashita~\cite{suzuki99siam}, for two robots executing deterministic
code and initially located at distinct positions. Our proof builds
upon the previously initiated Pactole framework~\cite{auger13sss} to use actual real numbers as locations instead
of rational numbers, and refines the definitions of executions
(including scheduling assumptions) to enable the study of executions
that involve only correct processes. Not only do we obtain a certified proof of the original impossibility result of Suzuki and Yamashita, we also get the more general impossibility result with an even number of robots, when any two robots are possibly initially at the same exact location.

\section{Preliminaries}

\subsection{Certification and the \coq proof assistant}\label{sec:coq}
To certify results and to guarantee the soundness of theorems, we use
the \coq proof assistant, a Curry-Howard based interactive prover
enjoying a trustworthy kernel.
The Pactole formal model is thus developed in \coq's formal language, a very
expressive $\lambda$-calculus: the \emph{Calculus of Inductive
  Constructions} (CIC)~\cite{coquand90colog}.
In this (functional) language, datatypes, objects, algorithms,
theorems and proofs can be expressed in a unified way, as terms.
$\lambda$-abstraction is denoted \lstinline!fun~x:T~=>~t!, and
application is denoted \lstinline!t~u!.
Curry-Howard isomorphism associates proofs and programs, types and
logical propositions. 
Writing a proof of a theorem in this setting amounts to building
(interactively in most cases but with the help of tactics) a term the
type of which corresponds to the theorem statement.
As a term is indeed a \emph{proof} of its type, ensuring the soundness
of a proof thus simply consists in type-checking a $\lambda$-term.

\coq has already been successfully employed for various
tasks such as the formalisation of programming language
semantics~\cite{Leroy-backend,Mccarthy67correctnessof} or 
mathematical developments as involved as the
4-colours~\cite{Gonthier4color} or
Feit-Thompson~\cite{DBLP:conf/popl/Gonthier13} theorems.

The reader will find in~\cite{bertot04coqart} a very comprehensive overview
and good practices with reference to \coq.
Developing a proof in a proof assistant may nonetheless be tedious, or require
expertise from the user.
To make this task easier, Pactole proposes a formal model, as
well as lemmas and theorem, to specify and certify results about
networks of autonomous mobile robots.
It is designed to be robust and flexible enough to express most of the
variety of assumptions in robots network, for example with reference
to the considered space: discrete or continuous, bounded or
unbounded\ldots

We do not expect the reader to be an expert in \coq but of course the
specification of a model for mobile robots in \coq requires some
knowledge of the proof assistant. %
We want to emphasise 
that the framework eases the developer's task. %
The notations and definitions given hereafter should be read as the typed
functional expressions \mbox{they are.} 

The formal model we rely on, as introduced in~\cite{auger13sss}, exceeds
our needs 
with reference to Byzantine robots, which are
irrelevant in the present work. %
Thus, for the sake of readability, a few notations have been slightly
simplified
:  the pruned code essentially deals with taking into account
the empty set of Byzantine robots in demonic actions. 
The reader is invited to check that the actual code is almost
identical.

\subsection{The Formal Model}\label{sec:formal}
The Pactole model\footnote{Available at \url{http://pactole.lri.fr}} has been sketched
in~\cite{auger13sss} {to which we refer for further
  details}; we recall here its main characteristics.

Two important features of \coq are used: a formalism of \emph{higher-order}, which allows us to quantify over programs, demons, etc.,
and the possibility to define  \emph{inductive} and
\emph{coinductive} types~\cite{sangiorgi12book}, so as to express inductive and coinductive
datatypes and properties. 
Coinductive types are in particular of invaluable help to express
in a rather direct way infinite behaviours, infinite datatypes and
properties on them, as we shall see with demons.

Robots are anonymous, however we need to identify some of them in the
proofs. 
Thus, we consider given a finite set of \emph{identifiers}, isomorphic
to a segment of $\setN$. We omit this set \lstinline!G! (usually
inferred by \coq) unless it is necessary to characterise the number of robots.
If needed in the model, we can make sure that names are not used by
the embedded algorithm. 

Robots are distributed in space, at places called \emph{locations}.
We define a \emph{position} as a \emph{function} from a set of identifiers to
the space of locations. The set of locations we consider here is the
real line \setR.

Robots compute their target position from the observed configuration
of their siblings in the considered space. We also define permutations
of robots, that is bijective applications from
 $G$ 
to itself,
usually denoted hereafter by Greek letters.  Moreover, any correct
robot is supposed to act as any other correct robot in the same
context, that is, with the same 
 perception
of the environment.
For two real 
 numbers $k \neq 0$ and $t$, a \emph{similarity} is a
function mapping a location $x$ to $k\times (x - t)$, denoted
$\sem{k,t}$. Real 
 number $k$ is called the homothetic factor, and
$-k\times t$ is called the translation factor.
Similarities can be extended to positions, by applying the similarity
transform to the extracted location.
This operation will be (abusively) written 
 $\sem{k,t} (\gp)$. 
Similarities are used as transformations of frames of
reference.

For a robot $\robid_i$, a computation takes as an input an entire
position 
 $\gp$ 
as seen by $\robid_i$, in its own frame of reference (scale, origin,
etc.), and returns a real 
 number $l_i$
corresponding to a location (the \emph{destination point}) in the same
frame.  As the robots are \emph{oblivious} in the present context, the
scale factor is taken anew at each cycle. Moreover to avoid any
symmetry breaking mechanism based on identifiers, the result of $r$
must be invariant by permutation of robots. We call \emph{robograms}
the embedded computation algorithms that fulfil this fundamental
property.

Robograms may be naturally defined in a \emph{completely abstract
  manner}, without any concrete code, in our \coq model as follows.
\begin{lstlisting}
Record robogram := {
  algo : position -> location ;
  AlgoMorph : forall p q sigma, (q == p \o sigma^-1) -> algo p = algo q }.
\end{lstlisting}

Demonic actions 
consist of
a function associating to each correct robot a
real 
 number $k$ such that $k = 0$ and the
robot is not activated, or $k \neq 0$ and the robot is activated with
a scale factor.  An actual \emph{demon} is simply an infinite sequence
(stream) of demonic actions, that is a coinductive object.
\begin{lstlisting}
Record demonic_action := {frame : G -> R}.
CoInductive demon :=  NextDemon : demonic_action -> demon -> demon.
\end{lstlisting}

Characteristic properties of demons include \emph{fairness} and
synchronous aspects. We 
described in~\cite{auger13sss} how
fair, FSYNC, and SSYNC 
demons could be defined 
using %
 coinductive types. We 
show %
 in
Section~\ref{sec:preuve} how $k$-fair demons can be
expressed similarly.

Finally, an \emph{execution} $(\gp_i)_{i\in\mathbb{N}}$  from an initial position for (correct)
robots $\gp_0$ and a demon $(\mbox{\lstinline!locate_byz!}_i,
\mbox{\lstinline!frame!}_i)_{i\in \mathbb{N}},$ is an infinite
sequence such that
      \[\gp_{i+1}(x)=\left\lbrace\begin{array}{ll}
          r_{\sem{\mbox{\scriptsize\texttt{frame}}_i(x),gp_i(x)}}(\gp_i) 
          & \textrm{if } \mbox{\texttt{frame}}_i(x) \neq 0 \\
          \gp_i(x) 
          & \textrm{otherwise}\end{array}\right.\]

It is thus an object of type: 
\begin{lstlisting}
CoInductive execution := 
  NextExecution : (G -> location) -> execution -> execution.
\end{lstlisting}
Its computation is reflected by a corecursive function \lstinline!execute!.

\section{Certification of Impossibility}\label{sec:preuve}

The impossibility result we aim to prove formally is the following:
\begin{thm}\label{thm:impossible}
It is impossible to achieve the gathering of an even number of
oblivious robots moving on the real line \setR with  SSYNC $k$-fair
demons for all $k \geq 1$.
\end{thm}

In this section, we specialise and enrich the Pactole model to provide a formal proof of
this theorem. 
Note that for the sake of readability
some notations may be slightly simplified compared to the actual code,
available from \url{http://pactole.lri.fr}.

The main idea of the proof is taken from~\cite{suzuki99siam} while our
premises are different: we allow for an unbounded number of robots,
provided that it is even, and for an arbitrary initial position. On
the contrary, \cite{suzuki99siam} requires the initial position to
have robots at distinct locations.\footnote{This is why our results
  are not in contradiction with~\cite{suzuki99siam},~Theorem~3.4, that
  exhibits a solution for a number of robots $n\geq 3$.}

 To this goal:
\emph{(i)} we consider robots as points, that is two or more robots can occupy the
  same location, thus no constraint is added to the definition of a
  position, 
\emph{(ii)}
  we assume robots enjoy strong global multiplicity detectors, the same
  global position is thus used for the computations of all robots, 
\emph{(iii)} we consider that the travelling time is negligible, destination
  points returned by robograms are used directly to determine new
  locations, 
\emph{(iv)} we consider oblivious robots, that is a new frame is chosen by
  the demon for each activation of any robot, 
\emph{(v)} we take \lstinline!location! to be \lstinline!R!, the
  (axiomatic) definition of \setR in the \coq standard library
  \texttt{Reals}. 
  Note that we are considering an unbounded continuous
  space. 

\subsection{$k$-Fairness}\label{sec:kfair}
A demon is said to be \emph{$k$-fair} if it is fair and $k$-bounded,
that is such that
between two successive activations of any robot, all other robots can be
activated at most $k$ times. Roughly speaking, $k$-fairness expresses
the ratio between the most active robot and the less
active one, as well as  avoids the degenerated case of robots not being
activated.

Firstly we express the property that, for any two robots \lstinline!g!
and \lstinline!h!, the demon activates \lstinline!g! within the $k$
next activations of
\lstinline!h!. %
It consists in three cases of activation for an inital round. %
Either \lstinline!g! is activated (its new frame is non-null) and we
are done; this is case \lstinline!kReset!, a base case. %
Either \lstinline!g! is not activated but \lstinline!h! is, and the
property will holds for $k+1$ if it holds for $k$ for the remainder of
the demon 
(case \lstinline!kReduce!). %
Finally if none of the two considered robots is activated during this
round, the property holds for a certain $k$ if it holds in the
remainder of the demon (case \lstinline!kStall!) for the same $k$.
Notice that if the latter case happens indefinitely, then one cannot
prove \lstinline!Between g h d! since \lstinline!Between! is an \emph{inductive}
relation\footnote{The curly brackets around the first argument 
  (\texttt{\{G\}}) 
  set it as \emph{implicit}, which allows
  us to omit
  it later on. }. 
\begin{lstlisting}[language=Coq]
Inductive Between {G} g h (d : demon G) : nat -> Prop := 
| kReset : forall k, frame (demon_head d) g <> 0 -> Between g h d k 
| kReduce : forall k, frame (demon_head d) g = 0 
   -> frame (demon_head d) h <> 0 -> Between g h (demon_tail d) k 
   -> Between g h d (k + 1) 
| kStall : forall k, frame (demon_head d) g = 0 
   -> frame (demon_head d) h = 0 -> Between g h (demon_tail d) k 
   -> Between g h d k.
\end{lstlisting}
An infinite demon is thus $k$-fair, for a certain $k$, if \lstinline!Between! holds for
any couple of robots \emph{at any time}, that is if
the demon is $k$-fair (for the very same $k$) from the start and also
for the remainder of the demon. We can express this coinductive
property as follows.
\begin{lstlisting}[language=Coq]
CoInductive kFair {G} k (d : demon G) :=
  AlwayskFair : (forall g h, Between g h d k) -> kFair k (demon_tail d) 
   -> kFair k d.
\end{lstlisting}

Intended as a framework and a library, our formal development provides
several theorems about $k$-fairness that may prove useful, namely that
a $k$-fair demon is fair, that if a demon is $k$-fair, then it is
$k'$-fair for all $k'\geq k$, etc.

\subsection{Definition of Success}\label{sec:success}

A robogram is a solution to the Gathering problem
if robots reach the same, unknown beforehand, location within
finite time regardless of their initial positions.
First we define the property for a position \lstinline!pos! of having
all 
 robots at a same location 
\lstinline!pt!. %
\begin{lstlisting}[language=Coq]
Definition stacked_at {G} (pos : G -> location) (pt : location) := 
  forall r : G, pos r = pt.
\end{lstlisting}
Hence there is a gathering point for an execution at some step if for
all future execution steps, 
the location is the same for all robots. 
Such an infinite behaviour is a coinductive
property. 
\begin{lstlisting}[language=Coq]
CoInductive Gather {G} (pt : location) (e : execution G) := 
 Gathering : stacked_at (execution_head e) pt
  -> Gather pt (execution_tail e) -> Gather pt e.
\end{lstlisting}
This situation has to occur \emph{eventually}, which we thus define as
an inductive property.  
\begin{lstlisting}[language=Coq]
Inductive WillGather {G} (pt : location) (e : execution G) :=
  | Now : Gather pt e -> WillGather pt e
  | Later : WillGather pt (execution_tail e) -> WillGather pt e.
\end{lstlisting}
If this holds for a given robogram \lstinline!r! and a given demon
\lstinline!d! from any initial position then \lstinline!r! is a
solution to the Gathering problem for \lstinline!d!. 
\begin{lstlisting}[language=Coq]
Definition solGathering {G} (r : robogram G) (d : demon G) :=
  forall (p : G -> location), 
   exists pt : location, WillGather pt (execute r d p).
\end{lstlisting}

We will prove that with a well chosen demon, even as constrained as a
$k$-fair demon, there exists an execution
where robots are always apart (we prove that this notion is in
contradiction with being a solution). More precisely, there is an
execution that keep half the robots away from the other
half; that is: the position is split. 
In the following, \lstinline!(G uplus G)! denotes the union of two disjoint sets isomorphic to the same
segment of \setN, hence guaranteeing an even number of robots. By
construction, an element \lstinline!g! of the left (respectively
right) \lstinline!G! is denoted \lstinline!inl~g! (respectively
\lstinline!inr~g!). Moreover, recall that the location is obtained by
  application of the position (which is a function) to an identifier.

\begin{lstlisting}[language=Coq]
Definition Split {G} (p : (G uplus G) -> R) := 
  forall x y : G, p (inl x) <> p (inr y).
\end{lstlisting}
The following coinductive property characterises such an execution:
\begin{lstlisting}[language=Coq]
CoInductive Always_Split {G} (e : execution (G uplus G)) :=
  CAS : Split (execution_head e) 
   -> Always_Split (execution_tail e) 
   -> Always_Split e.
\end{lstlisting}
In fact, the faulty execution we exhibit with this property in the
proof leaves a particular position indefinitely \emph{bivalent}: with
the robots evenly distributed over two distinct locations only.

Of course, any execution for which this property holds cannot be
compatible with a solution for a non-empty set of robots (of even cardinality).
\begin{lstlisting}[language=Coq]
Theorem Always_Split_no_gathering : 
  forall (G : finite) (e : execution (G uplus G)),
   inhabited G -> Always_Split e -> forall pt, \not WillGather pt e.
\end{lstlisting}

\subsection{The Theorem in \coq}\label{sec:coqthm}

We may now state a formal version of Theorem~\ref{thm:impossible} as follows:
\begin{lstlisting}
Theorem noGathering : forall (G : finite) (r : robogram (G uplus G)),
  inhabited G
  -> forall k : nat, (1 <= k)
    -> \not (forall d, kFair k d -> solGathering r d)
\end{lstlisting} 
the proof of which amounts to showing that for a non-null even number
of robots, any \lstinline!k! and any robogram \lstinline!r! there
exists a $k$-fair demon that prevents $r$ to gather all robots.

The proof we formalise is inspired from~\cite{suzuki99siam}; it makes
use of two demons, one that is fully synchronous, and one that is
$1$-fair. Depending on the expected result of the first move, we use
one or the other.

We consider an initial position consisting of two separate piles of
the same number of robots.  If the expected first move brings the
robots of one pile onto the other pile, we choose the
fully-synchronous demon, which results in switching the locations of
the two piles, thus in obtaining an equivalent position. Otherwise, we
choose the $1$-fair demon that will activate only one pile at a time;
the piles moving alternatively, a change of frame suffices then to get
back to an equivalent position.

Both cases allow us to show that \lstinline!Always_Split! holds, thus
proving Theorem \lstinline!noGathering!.

\section{Remarks and Perspectives}\label{sec:concl}

Thanks to the abstraction level of the Pactole framework, setting the space to
be \setR, thus both unbounded and continuous, is not as complicated as
one could imagine; it emphasises the relevance of a formal proof
approach and how it is complementary to other formal verification
techniques. In addition to the
syntactical invocation of \lstinline!R! and associated functions, the
main change from previous formalisations (that in particular were
dealing with \setQ)
addresses proofs more than specifications, and lies in the fact that we
use axiomatic reals. With such a description of \setR, there is no
computation. Hence relations between two elements of type
\lstinline!R! must be actually proved as they usually cannot be
obtained by computation primitives.

The size of the specialised development for the relevant notions and
the aforementioned theorems (thus excluding for example the complete
library for reals) is quite small, as it is approximately 480 lines of specifications and
430 lines of proofs. The file \lstinline!noRDVevenR.v! itself is about
200 lines of specifications for 250 lines of proof scripts. This is a
good indication on how adequate our framework is,
as
proofs are not too intricate and remain human readable.

\bibliographystyle{plain}
\bibliography{../../../biblio}

\begin{thebibliography}{10}

\bibitem{auger13sss}
C\'edric Auger, Zohir Bouzid, Pierre Courtieu, S\'ebastien Tixeuil, and Xavier
  Urbain.
\newblock {Certified Impossibility Results for Byzantine-Tolerant Mobile
  Robots}.
\newblock In Teruo Higashino, Yoshiaki Katayama, Toshimitsu Masuzawa, Maria
  Potop-Butucaru, and Masafumi Yamashita, editors, {\em Stabilization, Safety,
  and Security of Distributed Systems - 15th International Symposium (SSS
  2013)}, volume 8255 of {\em Lecture Notes in Computer Science}, pages
  178--186, Osaka, Japan, November 2013. Springer-Verlag.

\bibitem{BMPTT13r}
B{\'e}atrice B{\'e}rard, Laure Millet, Maria Potop-Butucaru, Yann Thierry-Mieg,
  and S{\'e}bastien Tixeuil.
\newblock {Formal verification of Mobile Robot Protocols}.
\newblock Technical report, May 2013.

\bibitem{bertot04coqart}
Yves Bertot and Pierre Cast\'eran.
\newblock {\em Interactive Theorem Proving and Program Development. Coq'Art:
  The Calculus of Inductive Constructions}.
\newblock Texts in Theoretical Computer Science. Springer-Verlag, 2004.

\bibitem{bonnet12sss}
Fran\c{c}ois Bonnet, Xavier D\'{e}fago, Franck Petit, Maria Potop-Butucaru, and
  S\'{e}bastien Tixeuil.
\newblock {Brief Announcement: Discovering and Assessing Fine-grained Metrics
  in Robot Networks Protocols}.
\newblock In Richa and Scheideler \cite{sss12}, pages 282--284.

\bibitem{CFPS12j}
Mark Cieliebak, Paola Flocchini, Giuseppe Prencipe, and Nicola Santoro.
\newblock Distributed computing by mobile robots: Gathering.
\newblock {\em SIAM J. Comput.}, 41(4):829--879, 2012.

\bibitem{coquand90colog}
Thierry Coquand and Christine Paulin-Mohring.
\newblock {Inductively Defined Types}.
\newblock In Per Martin-L{\"o}f and Grigori Mints, editors, {\em {International
  Conference on Computer Logic ({C}olog'88)}}, volume 417 of {\em Lecture Notes
  in Computer Science}, pages 50--66. Springer-Verlag, 1990.

\bibitem{devismes12sss}
St\'{e}phane Devismes, Anissa Lamani, Franck Petit, Pascal Raymond, and
  S\'{e}bastien Tixeuil.
\newblock {Optimal Grid Exploration by Asynchronous Oblivious Robots}.
\newblock In Richa and Scheideler \cite{sss12}, pages 64--76.

\bibitem{flocchini12book}
Paola Flocchini, Giuseppe Prencipe, and Nicola Santoro.
\newblock {\em {Distributed Computing by Oblivious Mobile Robots}}.
\newblock Synthesis Lectures on Distributed Computing Theory. Morgan {\&}
  Claypool Publishers, 2012.

\bibitem{Gonthier4color}
Georges Gonthier.
\newblock {Formal Proof—The Four-Color Theorem}.
\newblock In {\em Notices of the AMS}, volume~55, page 1370. december 2008.

\bibitem{DBLP:conf/popl/Gonthier13}
Georges Gonthier.
\newblock {Engineering Mathematics: the Odd Order Theorem Proof}.
\newblock In Roberto Giacobazzi and Radhia Cousot, editors, {\em POPL}, pages
  1--2. ACM, 2013.

\bibitem{Leroy-backend}
Xavier Leroy.
\newblock {A Formally Verified Compiler Back-End}.
\newblock {\em Journal of Automated Reasoning}, 43(4):363--446, 2009.

\bibitem{Mccarthy67correctnessof}
John Mccarthy and James Painter.
\newblock {Correctness of a Compiler for Arithmetic Expressions}.
\newblock In {\em Proceedings of Applied Mathematica}, volume~19 of {\em
  Mathematical Aspects of Computer Science}, pages 33--41. American
  Mathematical Society, 1967.

\bibitem{sss12}
Andr{\'e}a~W. Richa and Christian Scheideler, editors.
\newblock {\em Stabilization, Safety, and Security of Distributed Systems -
  14th International Symposium (SSS 2012)}, volume 7596 of {\em Lecture Notes
  in Computer Science}, Toronto, Canada, October 2012. Springer-Verlag.

\bibitem{sangiorgi12book}
Davide Sangiorgi.
\newblock {\em {Introduction to Bisimulation and Coinduction}}.
\newblock Cambridge University Press, 2012.

\bibitem{suzuki99siam}
Ichiro Suzuki and Masafumi Yamashita.
\newblock {Distributed Anonymous Mobile Robots: Formation of Geometric
  Patterns}.
\newblock {\em SIAM Journal of Computing}, 28(4):1347--1363, 1999.

\end{thebibliography}

\end{document}